\documentclass[traditabstract, letter]{aa}
\usepackage{natbib}
\usepackage{epsfig}
\usepackage{txfonts}
\usepackage{graphicx}
\usepackage{float}

\usepackage{color,hyperref}
\definecolor{darkblue}{rgb}{0.0,0.0,0.4}
\hypersetup{colorlinks,breaklinks, linkcolor=darkblue,urlcolor=darkblue, anchorcolor=darkblue,citecolor=darkblue}

\def\ms{\,m\,s$^{-1}$}         
\def\kms{\,km\,s$^{-1}$}         
\def\ms{\hbox{\,m\,s$^{-1}\,$}}         
\def\m2s2{\hbox{\,m$^{2}$\,s$^{-2}$}} 
\def\kms{\hbox{\,km\,s$^{-1}$}}       
\def\vsini{\hbox{$\upsilon \sin i_{\star}\;$}}      
\def\Msun{\hbox{$\mathrm{M}_{\odot}$}}             
\def\Rsun{\hbox{$\mathrm{R}_{\odot}$}}
\def\Mjup{\hbox{$\mathrm{M}_{\rm Jup}$}}
\def\Rjup{\hbox{$\mathrm{R}_{\rm Jup}$}}

\def\teff{T$_{\rm eff}~$}

\bibpunct{(}{)}{;}{a}{}{,}
\defcitealias{2011AJ....142..195S}{Sh11}
\defcitealias{2012A&A...541A..56M}{Ma12}
\defcitealias{2011ApJ...736L...4S}{Sz11}
\defcitealias{2012MNRAS.421L.122S}{Szab{\'o} et al. 2012}
\defcitealias{2012MNRAS.422.1512M}{M\&H12}
\defcitealias{2011ApJS..197...10B}{Ba11}

\begin{document}
\title{\textsc{SOPHIE} velocimetry of \textit{Kepler} transit candidates \thanks{Based on observations made with SOPHIE on the 1.93-m telescope at Observatoire de Haute-Provence (CNRS), France}\fnmsep\thanks{RV data is only available at the CDS via anonymous ftp to cdsarc.u-strasbg.fr (130.79.128.5) or via http://cdsarc. u- strasbg.fr/viz- bin/qcat?J/A+A/vol/page}
}
\subtitle{VI. An additional companion in the KOI-13 system}

\author{
A. Santerne \inst{1,3} 
\and C. Moutou \inst{1}
\and S.~C.~C. Barros \inst{1}
\and C. Damiani \inst{1}
\and R.~F. D\'iaz \inst{1}
\and J.-M. Almenara \inst{1}
\and A.~S. Bonomo \inst{1}
\and F. Bouchy \inst{2,3}
\and M. Deleuil \inst{1}
\and G. H\'ebrard \inst{2,3}
}
\institute{
Laboratoire d'Astrophysique de Marseille, Universit\'e d'Aix-Marseille \& CNRS, UMR7326, 38 rue F. Joliot-Curie, 13388 Marseille Cedex 13, France 
\and Institut d'Astrophysique de Paris, UMR7095 CNRS, Universit\'e Pierre \& Marie Curie, 98bis boulevard Arago, 75014 Paris, France
\and Observatoire de Haute-Provence, Universit\'e d'Aix-Marseille \& CNRS, 04870 Saint Michel l'Observatoire, France
}
\date{Received  26 June 2012 / Accepted 6 July 2012}

\offprints{Alexandre~Santerne\\
 \email{alexandre.santerne@oamp.fr}}

\abstract{We report the discovery of a new stellar companion in the KOI-13 system. KOI-13 is composed of two fast-rotating A-type stars of similar magnitude. One of these two stars hosts a transiting planet discovered by \textit{Kepler}. We obtained new radial velocity measurements using the SOPHIE spectrograph at the Observatoire de Haute-Provence that reveal an additional companion in this system. This companion has a mass of between 0.4 \Msun\ and 1 \Msun\ and orbits one of the two main stars with a period of 65.831 $\pm$ 0.029 days and an eccentricity of 0.52 $\pm$ 0.02. The radial velocities of the two stars are derived using a model of two fast-rotating line profiles. From the residuals, we find a hint of the stellar variations seen in the \textit{Kepler} light curve, which have an amplitude of about 1.41\kms\ and a period close to the rotational period. This signal appears to be about three orders of magnitude larger than expected for stellar activity.
From the analysis of the residuals, we also put a 3-$\sigma$ upper limit on the mass of the transiting planet KOI-13.01 of 14.8 \Mjup\ and 9.4 \Mjup, depending on which star hosts the transit. We find that this new companion has no significant impact on the photometric determination of the mass of KOI-13.01 but is expected to affect precise infrared photometry. Finally, using dynamical simulations, we infer that the new companion is orbiting around KOI-13B and that the transiting planet candidate is expected to orbit KOI-13A. Thus, the transiting planet candidate KOI-13.01 is orbiting the main component of a hierarchical triple system.

}

\titlerunning{An additional companion in the KOI-13 system}
\authorrunning{A.~Santerne et al.}

\maketitle

\section{Introduction}
\label{intro}

Among the 2321 \textit{Kepler} Objects of Interests (KOI) discovered so far \citep{2012arXiv1202.5852B}, KOI-13 is one of the most interesting. It is a common proper-motion binary with two bright \citep[V$_{A} = 9.9$ mag and V$_{B} = 10.2$ mag,][]{1994CoORB.115....1D} stars angularly separated by $\sim$1.2\arcsec\ \citep{1904LicOB...3....6A}. One of these stars hosts a transiting planet candidate, KOI-13.01, orbiting every 1.76 days \citep{2012arXiv1202.5852B}. A first analysis performed by \citet{2011ApJ...736L...4S} \citepalias[hereafter][]{2011ApJ...736L...4S} determined that the planetary candidate is transiting the brightest component of the binary, KOI-13A. They also estimated the stellar parameters of both stars. They found that both the A and B components are A5-type stars with values of \vsini of $65\pm10$\kms\ and $70\pm10$\kms, respectively. From the analysis of the transit, they determined the radius of KOI-13.01 to be $2.2\pm0.1$\Rjup\ and concluded that it is an inflated brown dwarf. They also found an asymmetry in the transit light curve that is characteristic of a misaligned companion transiting a hot and fast-rotating star affected by the gravity darkening effect \citep[see][\citepalias{2011ApJS..197...10B} and references therein]{2011ApJS..197...10B}. By considering either beaming or ellipsoidal effect \citep{2011MNRAS.415.3921F}, several studies have tried to measure the mass of the transiting candidate using only the exquisite \textit{Kepler} photometry. \citet{2011AJ....142..195S} \citepalias[hereafter][]{2011AJ....142..195S} and \citet{2012A&A...541A..56M} \citepalias[hereafter][]{2012A&A...541A..56M} found a mass of $9.2\pm1.1$ \Mjup\ and $10\pm2$ \Mjup\ using the beaming effect, and \citet{2012MNRAS.422.1512M} \citepalias[hereafter][]{2012MNRAS.422.1512M}found a mass of $8.3\pm1.25$ \Mjup\ using the ellipsoidal effect only. Thus, KOI-13.01 appears to be a massive and inflated hot jupiter orbiting one component of a binary system. In this letter, we present the results of our radial velocity follow-up observations of this system.



\section{SOPHIE observations}
\label{sophieobs}

\subsection{Observations}

We performed spectroscopic observations of the KOI-13 system with the SOPHIE spectrograph \citep{2008SPIE.7014E..17P, 2009A&A...505..853B} mounted on the 1.93-m telescope at Observatoire de Haute-Provence, France. We acquired 17 high resolution spectra from 2011, March 25 to 2012, May 23\footnote{prog. IDs: 11A.PNP.MOUT, 11B.PNP.MOUT, 12A.PNP.MOUT} using the high efficiency mode ($R \sim 39~000$ at 550nm) of SOPHIE and the fast CCD read-out mode. The spectra were reduced with the online standard pipeline and the weighted cross-correlation functions (hereafter CCF) were computed using a F0 mask \citep{1996A&AS..119..373B, 2002A&A...388..632P}.

\subsection{Cross-correlation function modeling}

Since the two main stars of the KOI-13 system are separated by $\sim1.2\arcsec$, their fluxes were observed simultaneously within the $3\arcsec$ diameter of the SOPHIE fiber. Thus, the spectra of both stars are blended, and there is no straightforward method for identifying the two contributions. To measure the radial velocity (hereafter RV) of both stars in the KOI-13 system, we first normalized the CCF from the blaze response of the spectrograph. The normalized CCFs are plotted in Fig. \ref{2CCFmodel}. We then computed a rotational profile to fit the data. The observed profile of a rapidly rotating star is the result of the convolution of a rotational profile $G(\upsilon)$ with the instrumental profile \citep{2005oasp.book.....G}. For the high efficiency mode of SOPHIE, we used a Gaussian with a FWHM of 10\kms\ as the instrumental profile. The rotational profile $G(\upsilon)$ was defined to be \citep[][p. 465]{2005oasp.book.....G}

\begin{equation}
\label{RotProf}
G(\upsilon) = \frac{2\left(1 - \varepsilon\right)\sqrt{1-\left(\upsilon/\upsilon_{L}\right)^{2}}+\frac{1}{2}\pi\varepsilon\left[1-\left(\upsilon/\upsilon_{L}\right)^{2}\right]}{\pi\upsilon_{L}\left(1-\varepsilon/3\right)},
\end{equation}
where $\upsilon_{L}$ is identical to the sky-projected rotational velocity \vsini and $\varepsilon$ is a limb-darkening coefficient. For the CCF modeling, we fixed the limb-darkening coefficient to $\varepsilon = 0.3$ \citep[see][Fig. 17.6]{2005oasp.book.....G}. 

\subsection{Distinguishing the radial velocities of both components}
\label{Disentangling}

\begin{figure}[]
\begin{center}
\includegraphics[width=\columnwidth]{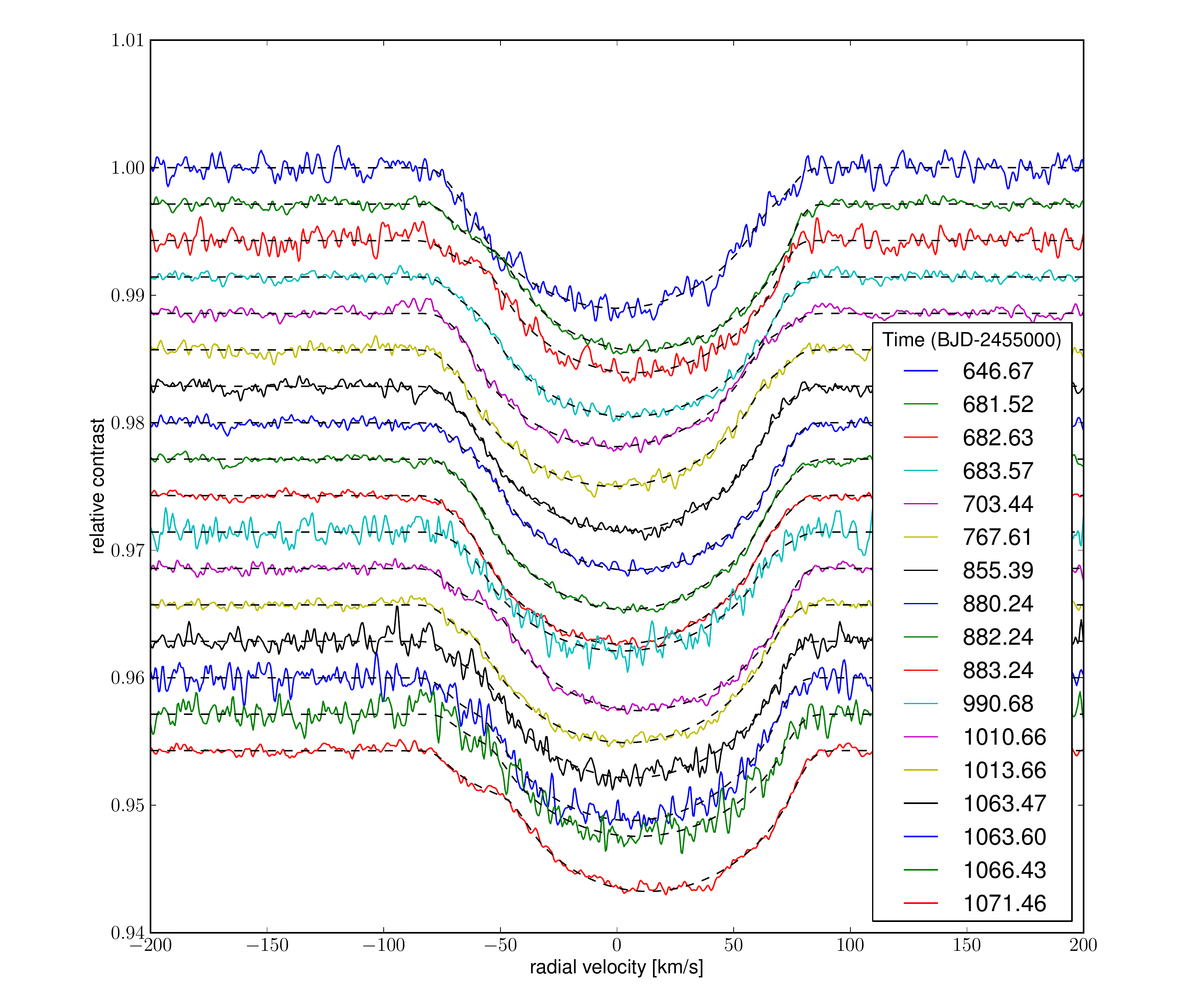} \\
\vspace{-0.2cm}
\includegraphics[width=\columnwidth]{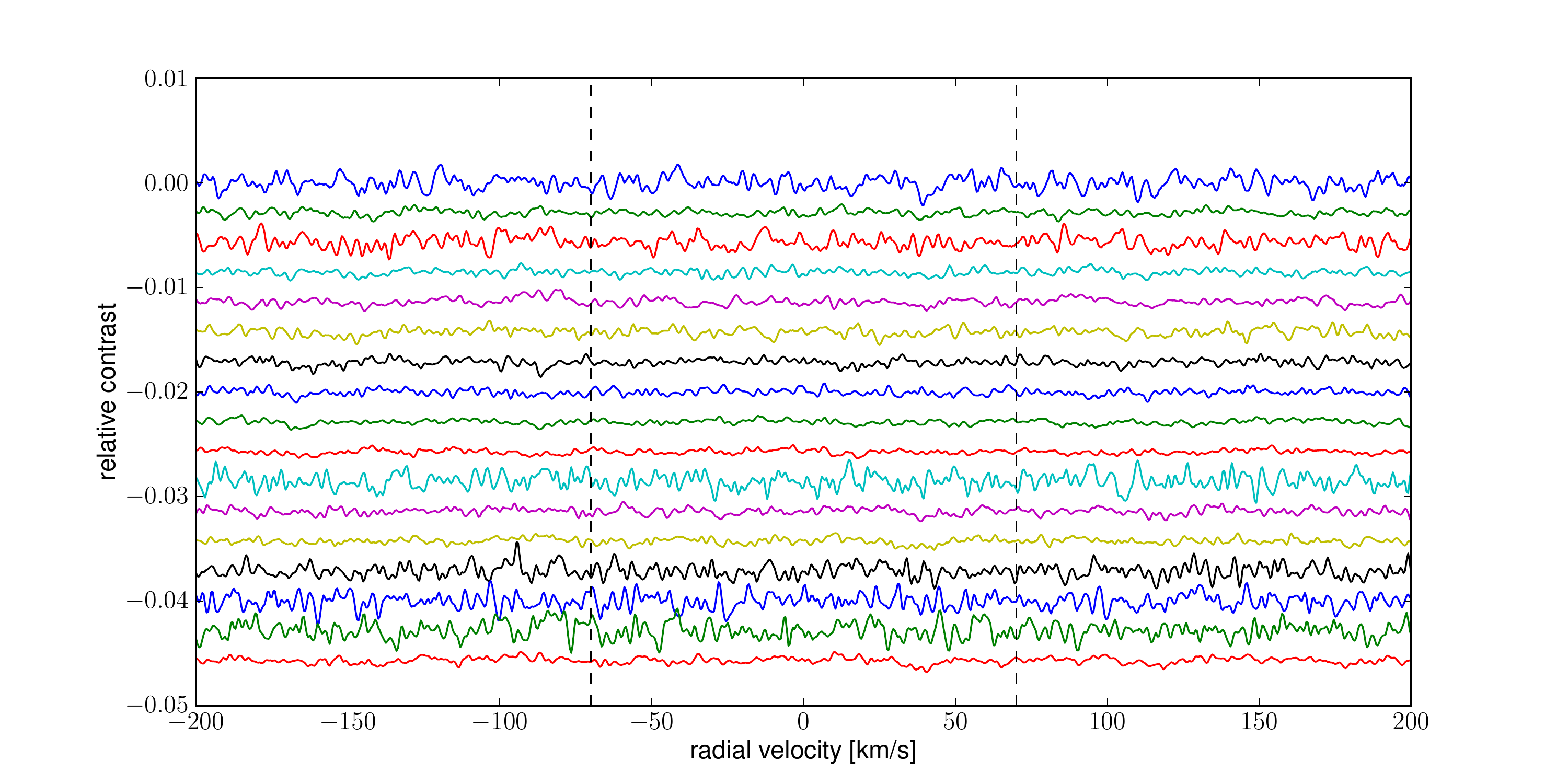}
\caption{Upper panel: Individual CCFs superimposed on a two-star CCF model (black lines).The  CCFs are arbitrary shifted down with increasing time. Lower panel: The CCF residuals displayed in the same way as in the upper panel. The two vertical lines indicate the limits of the CCFs.}
\label{2CCFmodel}
\end{center}
\end{figure}

To adjust the CCFs, we thus constructed a model with two fast-rotating stars. We fitted the 17 CCFs simultaneously with this model of two fast-rotating stars allowing both the RV, contrasts and \vsini to vary. We adjusted the contrast of the two stars for each CCF independently since both the seeing and the telescope guiding may favor one or the other star during the exposure. The best-fit model of each CCF as well as the residuals are displayed in Fig. \ref{2CCFmodel}. One of the two stars seems to have a CCF profile with a larger area. We know from \citet{1986MNRAS.223..581H} that the relative profile area of a binary can be used to measure its luminosity ratio. However, this is not enough evidence to be able to match each profile with the known star. In consequence, we refer to the component with a slightly larger (smaller) profile area as KOI-13$\alpha$ (KOI-13$\beta$, respectively) to avoid any confusion. We find $\vsini_{\alpha} = 76.6 \pm 0.2$\kms\ and $\vsini_{\beta} = 62.7 \pm 0.2$\kms. We note that these values agree with the values found by \citetalias{2011ApJ...736L...4S} to within 1-$\sigma$. The median values for the contrasts are $c_{\alpha} = 0.58 \pm 0.09 \%$ and $c_{\beta} = 0.59 \pm 0.05 \%$.

\begin{figure*}[]
\begin{center}
\includegraphics[width=1.7\columnwidth]{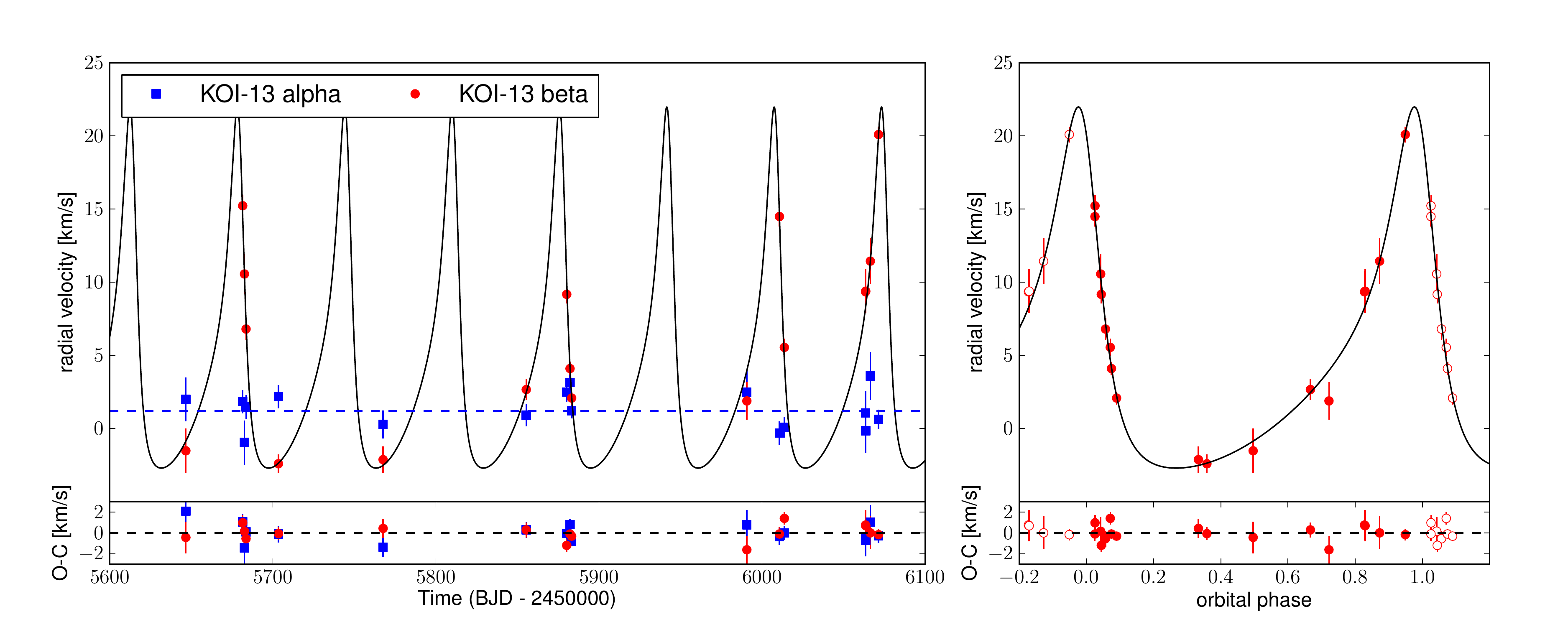}
\caption{Radial velocity curves and residuals of KOI-13$\alpha$ (blue squares) and KOI-$\beta$ (red circles) measured by SOPHIE against time (left panels) and phase-folded to the best-fit period of KOI-13$\beta$ (right panels). The dashed blue line represents the systemic RV of KOI-13$\alpha$.}
\label{RVtime}
\end{center}
\vspace{-0.8cm}
\end{figure*}

The derived RV and their uncertainties for both stars are listed in Table \ref{RVs} and displayed in Fig. \ref{RVtime}. We found no clear RV variation for the star KOI-13$\alpha$ with a systemic RV of $1.20 \pm 1.24$\kms. On the other hand, KOI-13$\beta$ has a large amplitude RV variation. We performed a Keplerian fit to the data of KOI-13$\beta$ using a Levenberg-Marquardt algorithm. The best solution has an eccentricity of 0.52 $\pm$ 0.02, a period of 65.831 $\pm$ 0.029 days, and a semi-amplitude of 12.42 $\pm$ 0.42 \kms. The parameters are listed in Table \ref{KOI-13params} with their uncertainties. This variation is caused by the reflex motion of the star orbited by a stellar companion with a minimum mass of $ 0.40 \pm 0.03 $ \Msun. We called this new stellar companion in the system KOI-13$\gamma$. We estimated a detection limit for a third component in the CCFs, after removing the main contributions of both A5 stars. The full set of SOPHIE spectra could exclude the presence of a third Gaussian non-rotating profile with an amplitude of 0.4\%, with more than 10-$\sigma$ confidence. Within the first-order assumption that the flux ratio scales as the CCF area, one can derive an upper-limit to the mass of KOI-13$\gamma$ of 1 \Msun. With less conservative assumptions, this upper limit decreases to 0.7 \Msun\ -- when 11 of the 17 residual CCFs can exclude a third component at the 3-$\sigma$ limit. In the following, we choose the conservative approach and an upper limit of 1 \Msun. KOI-13$\gamma$ is not transiting its host star since a deep transit would have been seen in the \textit{Kepler} light curve. Considering the whole possible mass range for KOI-13$\gamma$ and the expected radii \citep{1996ima..book.....C}, we estimated that an orbital inclination of up to 87$^{\circ}$ is compatible with the absence of a transit \citep[eq. 7,][]{2010arXiv1001.2010W}. Using eq. 1 from \citet{2011MNRAS.415.3921F}, we estimated that KOI-13$\gamma$ should produce a beaming effect on KOI-13$\beta$ at the level of about 165ppm, without any dilution. This amplitude may be detectable with a careful filtering of the \textit{Kepler} light curve that is beyond the scope of this letter.

\onltab{1}{
\begin{table*}[p]
  \centering 
  \caption{Radial velocities of KOI-13$\alpha$ and KOI-13$\beta$.}\label{RVs}
\setlength{\tabcolsep}{4mm}
\begin{minipage}[c]{2\columnwidth} 
\centering
\begin{tabular}{ccccccc}
\hline
BJD & RV$_{\alpha}$ & $\pm1\sigma_{\alpha}$ & RV$_{\beta}$ & $\pm1\sigma_{\beta}$ & Texp & SNR/pix\\
 - 2450000 & [\kms] & [\kms] & [\kms] & [\kms] & [s] & @550nm\\
\hline
5646.66816  &  1.992  &  1.499  &  -1.518  &  1.520  &  586  &  44.7\\
5681.51999  &  1.837  &  0.793  &  15.224  &  0.742  &  1146  &  100.9\\
5682.62863  &  -0.956  &  1.514  &  10.562  &  1.356  &  358  &  49.5\\
5683.57086  &  1.479  &  0.804  &  6.802  &  0.757  &  930  &  95.6\\
5703.44229  &  2.183  &  0.795  &  -2.405  &  0.632  &  1800  &  115.7\\
5767.61032  &  0.281  &  0.946  &  -2.123  &  0.907  &  1800  &  77.0\\
5855.38789  &  0.900  &  0.760  &  2.666  &  0.717  &  2700  &  100.3\\
5880.23849  &  2.494  &  0.669  &  9.172  &  0.632  &  3600  &  120.8\\
5882.23914  &  3.144  &  0.506  &  4.099  &  0.455  &  3600  &  186.9\\
5883.23991  &  1.203  &  0.503  &  2.086  &  0.441  &  3600  &  196.8\\
5990.68207  &  2.483  &  1.424  &  1.888  &  1.291  &  3600  &  52.1\\
6010.66131  &  -0.308  &  0.813  &  14.487  &  0.673  &  1845  &  110.7\\
6013.65902  &  0.085  &  0.696  &  5.547  &  0.605  &  2328  &  125.6\\
6063.47003  &  1.062  &  1.493  &  9.345  &  1.443  &  981  &  46.6\\
6063.59547  &  -0.155  &  1.527  &  9.390  &  1.492  &  1201  &  45.0\\
6066.43484  &  3.586  &  1.638  &  11.441  &  1.582  &  1669  &  42.5\\
6071.46067  &  0.629  &  0.668  &  20.092  &  0.540  &  3600  &  152.3\\
\hline
\end{tabular}
\end{minipage}
\end{table*}
}

\begin{table}[]
  \centering 
  \caption{KOI-13$\gamma$ orbital parameters}
  \label{KOI-13params}
\begin{minipage}[c]{\columnwidth} 
\renewcommand{\footnoterule}{}                          
\begin{tabular}{lc}
\hline
Orbital period $P$ [d] & 65.831 $\pm$ 0.029 \\
Epoch of periastron T$_\mathrm{p}$ [BJD-245000] & 5943.155 $\pm$ 0.181 \\
Eccentricity $e$ & 0.52 $\pm$ 0.02 \\
Argument of periastron $\omega$ [$^{\circ}$] & 32 $\pm$ 3 \\
Systemic radial velocity $\upsilon_{0}$ [\kms] & 3.98 $\pm$ 0.24  \\
Radial velocity semi-amplitude $K$ [\kms] & 12.42 $\pm$ 0.42 \\
\hline
\end{tabular}
\end{minipage}
\end{table}

\subsection{Stellar variability}

Another signal in the \textit{Kepler} light curve varies with a semi-amplitude of about $\sim$20 ppm and a period of $1.059 \pmÊ0.002$ d \citepalias{2012A&A...541A..56M,2011AJ....142..195S, 2012MNRAS.421L.122S}. The discoveres of this signal claimed that the signal has a stellar origin. \citet{2012MNRAS.421L.122S} also claimed that the signal has a period close to the expected rotational period of the star and may be either granulation noise or starspot features \citep[see also][]{2011MNRAS.415.1691B}. 

\begin{figure}[h]
\begin{center}
\setlength{\tabcolsep}{0.mm}
\includegraphics[width=0.65\columnwidth]{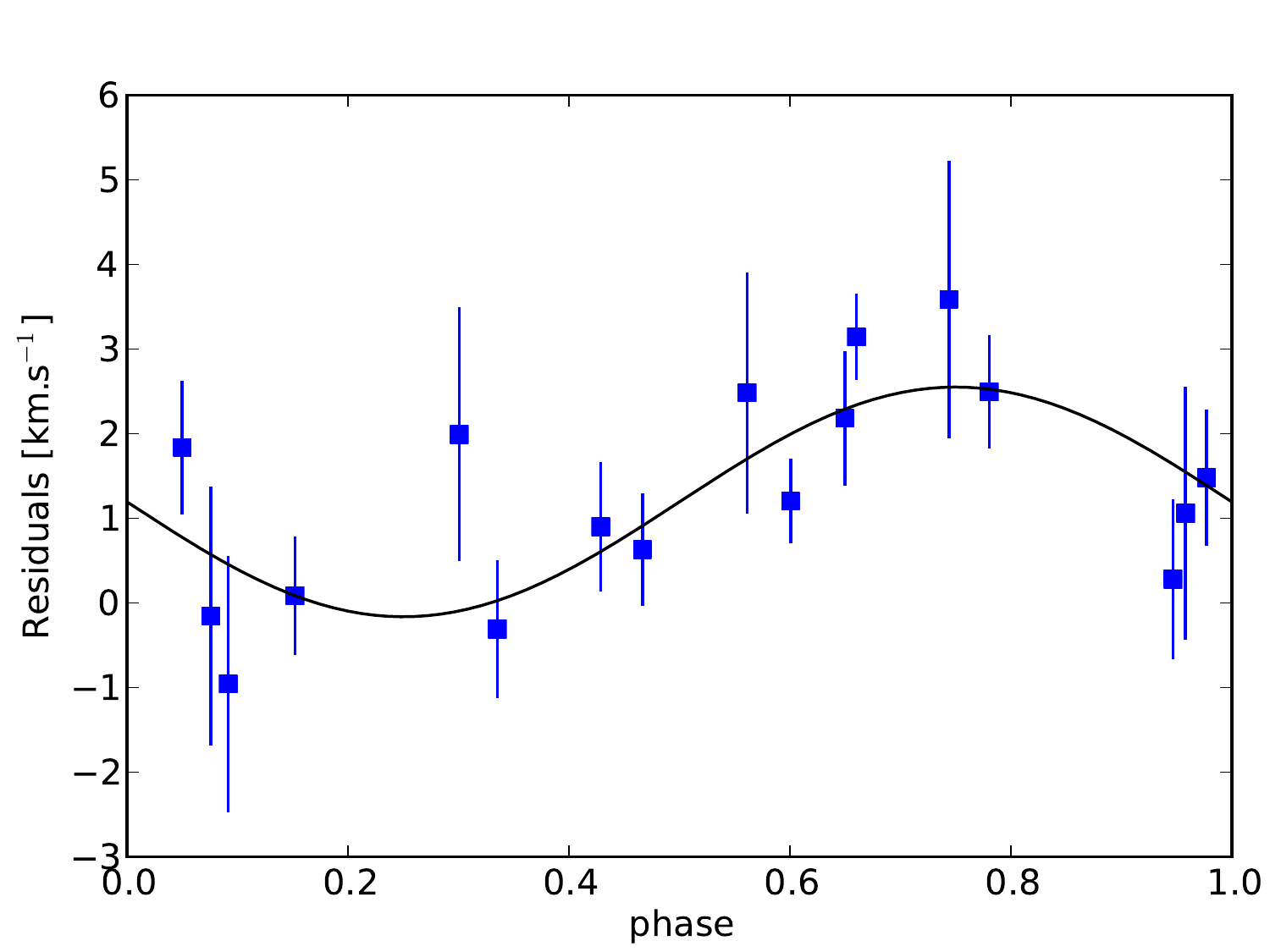}
\caption{Radial velocity residuals of KOI-13$\alpha$ phase-folded to the best-solution period (P = 1.0642 d). The black line is the best-fit circular model that fits the data.}
\label{Prot}
\end{center}
\end{figure}

We phase-folded the residuals of both stars to the expected stellar-rotational period within the errors. We found that the residuals of KOI-13$\alpha$ appear to have a circular variation with a semi-amplitude of $1.41 \pmÊ0.38$ \kms\ (3.6-$\sigma$ significance level). The best-fit period of this signal detected by RV is $1.0642 \pm 0.0003$ d, at 2.6-$\sigma$ from the photometric period. This amplitude is slightly larger than the expected variation caused by the planet candidate of about 1\kms. The residuals of KOI-13$\alpha$, phase-folded at the stellar rotational period are plotted in Fig. \ref{Prot}. No similar signal is seen in the residuals of KOI-13$\beta$. If the variation seen in the light curve is due to spots that produce a depth of about 60ppm, we might have expected to measure a RV amplitude at the level of a few \ms \citep{2012MNRAS.419.3147A}. If the signal seen in RV is real, it should have a different origin from starspots and might represent stellar pulsation \citepalias{2011AJ....142..195S}, as in the case of WASP-33 \citep{2010MNRAS.407..507C}.

\section{On the mass of the \textit{Kepler} transiting candidate KOI-13.01}

\subsection{Radial velocity upper-limit}

\begin{figure}[]
\begin{center}
\setlength{\tabcolsep}{0.mm}
\begin{tabular}{cc}
\includegraphics[width=0.5\columnwidth]{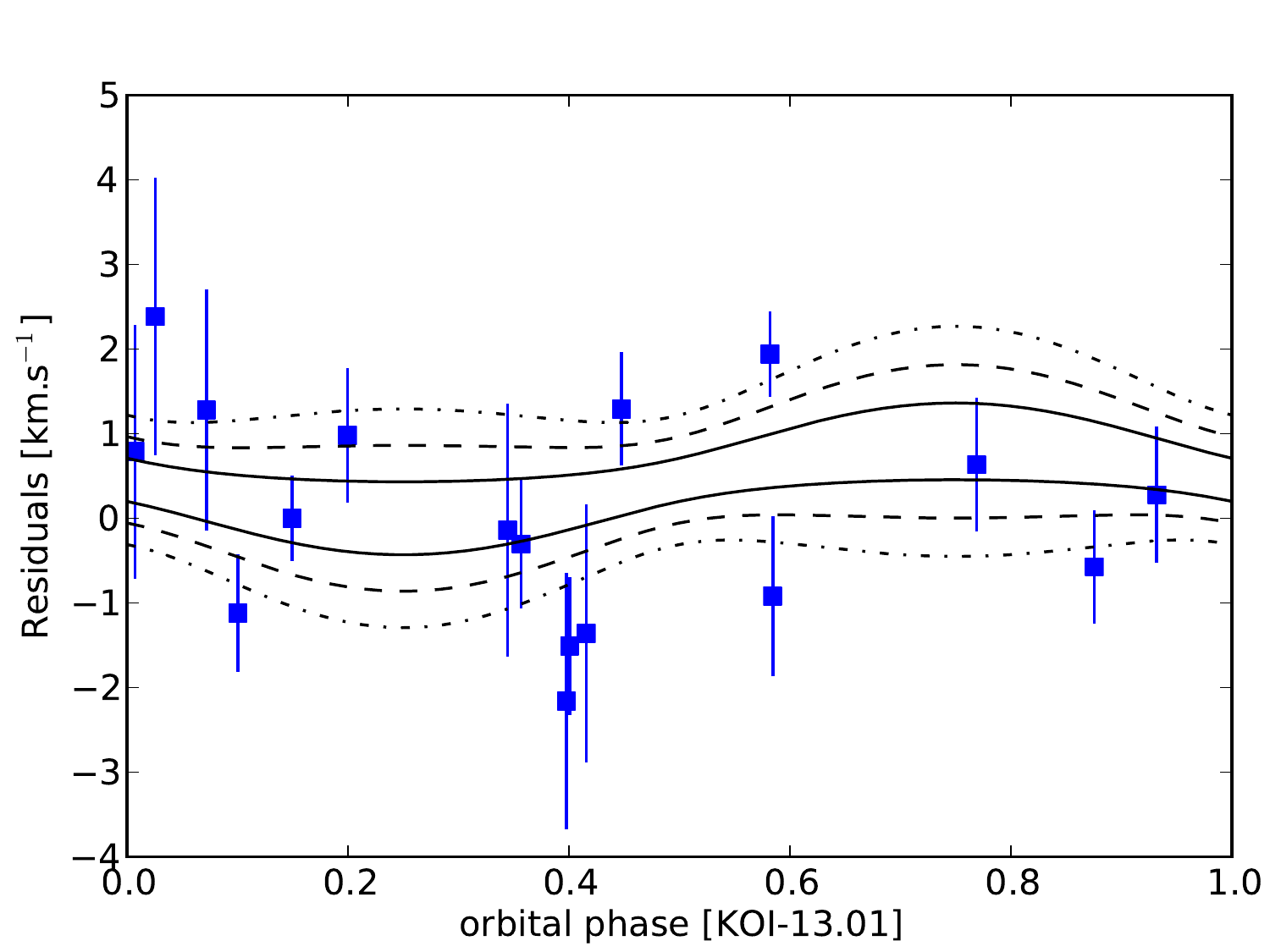} &  \includegraphics[width=0.5\columnwidth]{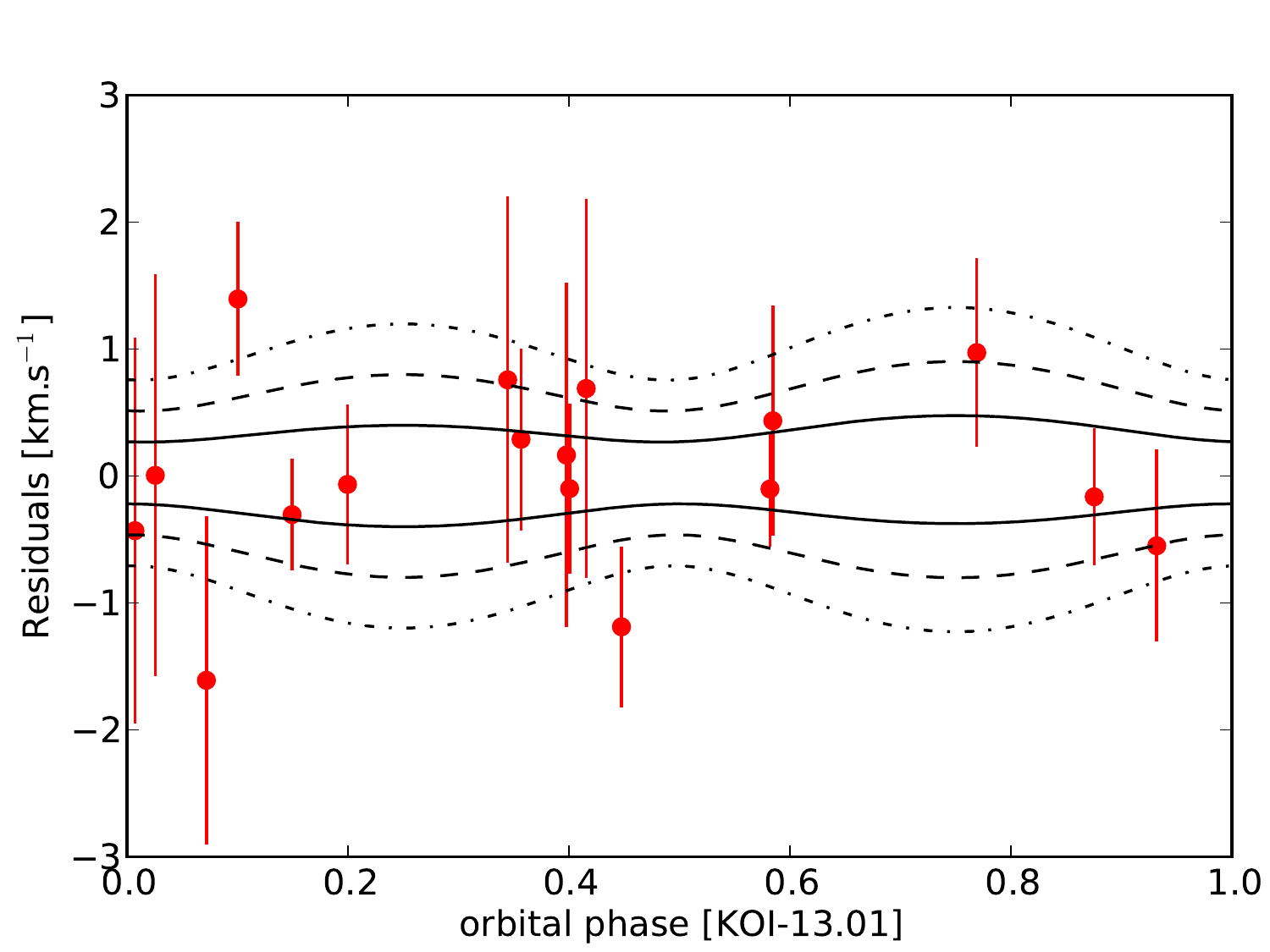}\\
\end{tabular}
\caption{Radial velocity residuals of KOI-13$\alpha$ (left panel) and KOI-13$\beta$ (right panel) phase-folded to the ephemeris of the \textit{Kepler} transiting-planet candidate KOI-13.01 (P$\sim$ 1.76 d). The solid, dashed, and dot-dashed lines represent the 1-$\sigma$, 2-$\sigma$, and 3-$\sigma$ (respectively) contours of a circular orbit.}
\label{RVph}
\end{center}
\vspace{-0.8cm}
\end{figure}

We used the residuals of the two components $\alpha$ and $\beta$ to place constrains on the mass of KOI-13.01 detected by \textit{Kepler}. We phase-folded the residuals (see Fig. \ref{RVph}) at the \textit{Kepler} ephemeris (P=1.763589 d, t$_{0}$=54953.56498 BJD). The 3-$\sigma$ upper limit is 14.8 \Mjup\ or 9.4 \Mjup, depending on whether the host star is KOI-13$\alpha$ or KOI-13$\beta$, respectively. We note that to be conservative, we did not remove the stellar variability from KOI-13$\alpha$. Removing this signal, we found a slightly lower upper-limit to the mass of KOI-13.01 of 14.4 \Mjup. The asymmetric transits of KOI-13.01 shown by \citetalias{2011ApJ...736L...4S} and \citetalias{2011ApJS..197...10B} must be produced by a hot and fast rotating star, completely ruling out KOI-13$\gamma$  as the host of the transiting planet candidate. If KOI-13.01 were transiting KOI-13$\alpha$, our results would agree with the mass derived using only photometry by \citetalias{2011AJ....142..195S}, \citetalias{2012A&A...541A..56M}, and \citetalias{2012MNRAS.422.1512M} and would confirm the planetary nature of KOI-13.01. On the basis of an upper limit of 9.4Ê\Mjup\ at 3-$\sigma$ compared to the photometric determinations, the scenario in which KOI-13.01 is transiting KOI-13$\beta$ seems more unlikely.

\subsection{Impact of KOI-13$\gamma$ on the photometric mass determination of KOI-13.01}

The mass estimate of KOI-13.01 was performed using the amplitude of the beaming or ellipsoidal effects \citepalias{2011AJ....142..195S, 2012A&A...541A..56M, 2012MNRAS.422.1512M}. This photometric determination does not take into account the amount of light from KOI-13$\gamma$. Equations 6 and 7 of the beaming and the ellipsoidal effects provided by \citetalias{2012A&A...541A..56M} show that the mass is proportional to the amplitude of these effects. The flux from KOI-13$\gamma$ impacts linearly on the mass determination of KOI-13.01. 

We investigated the impact of this new stellar companion on the mass of the planetary companion. We computed the dilution factor that affects the light curve of KOI-13A, which is expected to be the host of the transiting planet \citepalias{2011ApJ...736L...4S}. We generated Planck curves of two A-type stars with radii of 2.55 \Rsun\ and 2.38 \Rsun\ and \teff of 8511 K and 8222 K, respectively \citepalias{2011ApJ...736L...4S}. We generated a third Planck curve for KOI-13$\gamma$, assuming a true mass, radius, and \teff according to table B.1 of \citet{1996ima..book.....C}. We integrated the three Planck curves within the \textit{Kepler} and \textit{Spitzer} IRAC at 3.6$\mu$m and 4.5$\mu$m bandpasses. We allowed the mass of KOI-13$\gamma$ to vary between 0.4 \Msun\ and 1 \Msun\ as its inclination is unknown, but constrained by SOPHIE spectra. We found that KOI-13$\gamma$ increases the dilution factor affecting KOI-13A by less than 0.8\% within the \textit{Kepler} bandpass. Thus, the photometric measurement of KOI-13.01's mass is insignificantly affected. However, in the infrared, the additional dilution factor produced by KOI-13$\gamma$ reaches up to 2.2\% for both sets of IRAC dara at 3.6$\mu$m and 4.5$\mu$m. Thus, this third star is expected to affect the precise characterization of the planetary atmosphere. 

\section{Constraining the scenarios}
\label{scenarios}
With an angular separation of 1.2\arcsec, the two main components of the KOI-13 system were observed simultaneously within the 3\arcsec\ diameter fiber of the SOPHIE spectrograph. We detected a new companion in the system, called KOI-13$\gamma$. We were unable to measure the mass of the transiting-planet candidate KOI-13.01 nor determine about which star, $\alpha$ or $\beta$, it orbits. \citet{2011ApJ...736L...4S} determined that the transiting planet candidate KOI-13.01 is transiting the star A using a ground-based lucky-imaging photometry. This suggests two different scenarios: 1) both companions orbit the same star, i.e. $\beta \equiv A$ and $\alpha \equiv B$ or, 2) both companions orbit a different star, i.e. $\alpha \equiv A$ and $\beta \equiv B$.
From the line profile area, we assumed that KOI-13$\alpha$ is the brightest component. The \vsini values derived from the CCF fit also hints that $\alpha \equiv A$ and $\beta \equiv B$. 

To more tightly constrain the two scenarios, we performed dynamical simulations using the \texttt{Mercury6} code \citep{1999MNRAS.304..793C}, assuming that KOI-13.01 and KOI-13$\gamma$ orbit the same star (scenario 1) and allowing different inclinations and masses for KOI-13$\gamma$. As the inclination of the orbital plane of KOI-13.01 \citepalias[85.9$^{\circ}$,][]{2011ApJS..197...10B} is of the same order as the maximum inclination of KOI-13$\gamma$ (87$^{\circ}$, see section \ref{Disentangling}), the orbits can be coplanar. In this configuration, the eccentricity of KOI-13.01 can be excited between 0 and 0.15 with periodical variations over about 30 years. The time of the occultation found in the \textit{Kepler} light curve \citepalias{2011ApJ...736L...4S} indicates that the orbit is highly circular \citepalias{2012A&A...541A..56M}. From our simulations, we found that a non-zero eccentricity will be detected within the lifetime of \textit{Kepler}. However, we also found that such a configuration would produce transit-time variations (TTV) at the level of several minutes. We analyzed the Q2 and Q3 short-cadence \textit{Kepler} light curve and found no TTV at the level of 24 seconds, at 3-$\sigma$, using a procedure similar to \citet{2011MNRAS.416.2593B}. If KOI-13$\gamma$ has a different inclination than KOI-13.01, the orbit of the latter is either unstable (for a mass of KOI-13$\gamma > 0.6$ \Msun) or should display significant TTV, non-zero eccentricity, as well as inclination variations on a short timescale. We thus conclude that KOI-13$\gamma$ and KOI-13.01 are orbiting two different stars, which confirms that $\alpha \equiv A$ and $\beta \equiv B$.

\section{Conclusion and discussions}

We have performed RV follow-up observations of the KOI-13 system with SOPHIE. The KOI-13 system is dominated by two bright A-type stars with \vsini of approximately 70\kms. By modeling the observed CCFs using two rotational profiles, we have been able to distinguish the RVs of the two stars and found that this system hosts an additional companion in the stellar regime, KOI-13$\gamma$. This companion has a mass of between 0.4 \Msun\ and 1 \Msun, an orbital period of 65.831 $\pm$ 0.029 days and an eccentricity of 0.52 $\pm$ 0.02. This previously unknown stellar companion does not significantly increase the dilution of the \textit{Kepler} light curve, thus does not affect the photometric mass determination of KOI-13.01. However, this companion is expected to make a significant contribution to infrared bandpasses. In the residuals of KOI-13$\alpha$, we found a hint of stellar variability with a period of about 1.064d, which is close to the period determined by photometry. This variation has an amplitude of $1.41 \pmÊ0.38$ \kms\ and is slightly longer than the expected variation due to the planet candidate KOI-13.01 ($K\lesssim1$\kms). If this signal is real, constraining the mass of KOI-13.01 using RV data will be challenging. We have also provided a 3-$\sigma$ upper-limit constraint based on the RV residuals of about 14.8 \Mjup\ or 9.4 \Mjup\ on the mass of KOI-13.01, depending on whether KOI-13$\alpha$ or KOI-13$\beta$ is transited, respectively. These limits in mass are compatible with the photometric measurements performed by \citetalias{2011AJ....142..195S}, \citetalias{2012A&A...541A..56M}, and \citetalias{2012MNRAS.422.1512M}. KOI-13.01 thus appears to be a massive transiting planet orbiting a member of a triple system. 

Finally, we used dynamical simulations of the KOI-13 system with the \texttt{Mercury6} code \citep{1999MNRAS.304..793C} to show that both companions cannot orbit the same star and therefore conclude that KOI-13$\alpha$ is KOI-13A and KOI-13$\beta$ is KOI-13B.

\begin{acknowledgements}
Financial support for the SOPHIE observations from the ÒProgramme National de Plan\'etologieÓ (PNP) of CNRS/INSU, France is gratefully acknowledged. We also acknowledge support from the French National Research Agency (ANR-08-JCJC-0102-01). R.F. D. is supported by CNES. A. S. thanks Coraline Dalibot for her contribution in developing the rotation profile model. We thank the anonymous referee for his/her fruitful comments. \end{acknowledgements}

\end{document}